# Surface-Illuminated Interdigitated Ge-on-Si Avalanche Photodiode With Separated Ge Absorption, Si Multiplication, and Lateral Field Engineering

Mason Clewis, Kuang-Yun Tu, Ching-Yu Hsu, and Zingway Pei, *Senior Member, IEEE*

**Abstract—This work presents a surface-illuminated lateral Ge-on-Si interdigitated separate-absorption-charge-multiplication (SACM) avalanche photodiode for 1550-nm detection. The intrinsic-region spacing, P⁻ charge-control width, and finger geometry are engineered to suppress high-field penetration toward the Ge/Si heterointerface and localize avalanche multiplication in Si. Simulations show that wider charge-control or intrinsic regions reduce fixed-bias dark current and shift avalanche onset to higher reverse bias. Among the tested structures, the intrinsic-region spacing/P⁻ charge-control width of 0.5/0.15 µm achieved a peak responsivity of 11.81 A/W at 1550 nm. At a −20 V applied voltage, it exhibited a multiplication gain of 3.796, a 3-dB bandwidth of 7.13 GHz, and a matched-bias gain-bandwidth product of 27.07 GHz. Optical-power-dependent measurements also show avalanche-gain compression consistent with space-charge screening. These results demonstrate effective lateral-field engineering for high-speed, surface-illuminated Ge-on-Si APDs.**

## I. INTRODUCTION

Optical interconnects are becoming increasingly important because they offer higher bandwidth, lower latency, and lower energy consumption than electrical interconnects, driven by the rapid growth of artificial intelligence, cloud computing, and the Internet of Things [1], [2].

This work was supported by the National Science Foundation (NSF) under Grant 2505427; the National Science and Technology Council (NSTC), Taiwan, by grants (NSTC 113-2221-E-005-037-MY3, NSTC-114-2640-E-005-002-, and NSTC-115-2640-E-005-002-); and the Ministry of Education through the Higher Education Sprout Project. (*Corresponding author: Zingway Pei.*)
Mason Clewis is with CREOL, The College of Optics and Photonics, University of Central Florida, Orlando, FL 32816 USA (e-mail: ma882633@ucf.edu).
Kuang-Yun Tu and Zingway Pei are with the Graduate Institute of Optoelectronic Engineering and the i-Center for Advanced Science and Technology (i-CAST), National Chung Hsing University, Taichung 40227, Taiwan (e-mail: tommy97121@smail.nchu.edu.tw; zingway@dragon.nchu.edu.tw).
Ching-Yu Hsu is with the College of Photonics, National Yang Ming Chiao Tung University, Hsinchu 30010, Taiwan (e-mail: good50612@gmail.com).

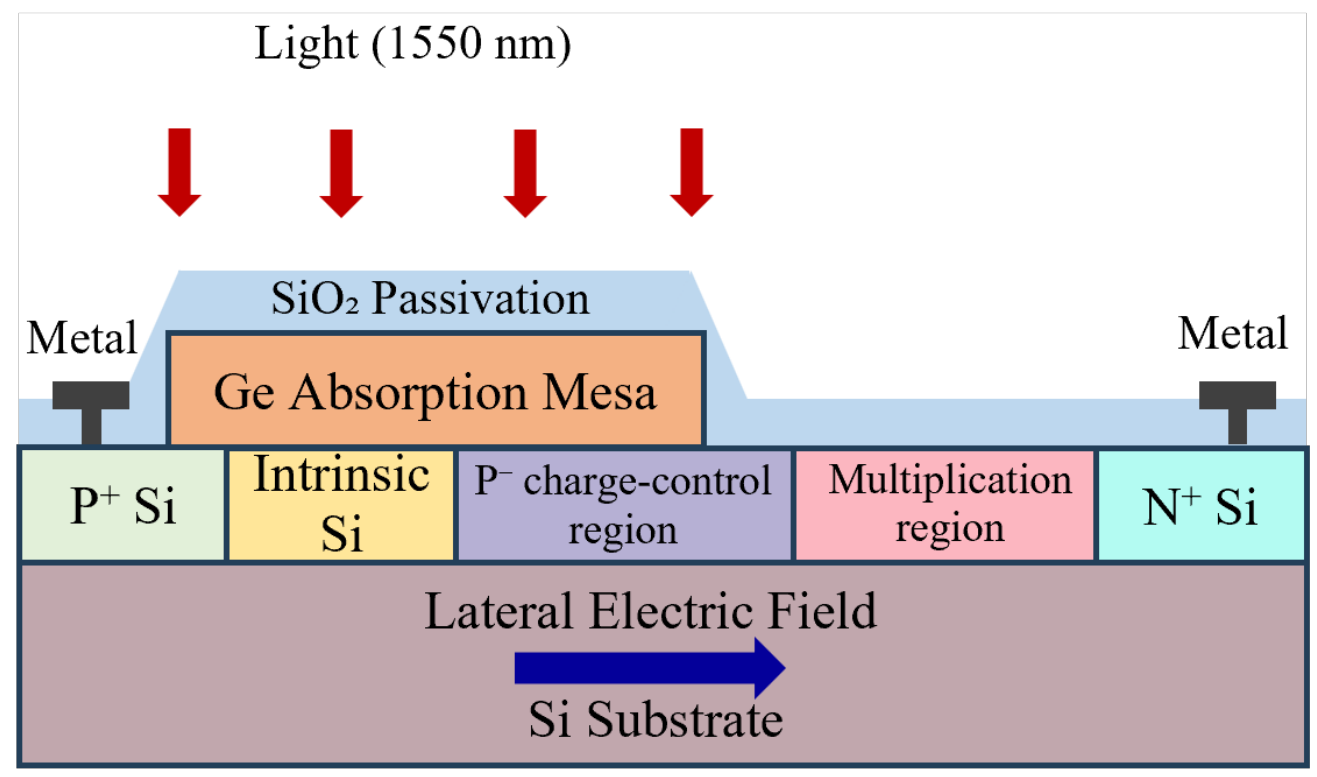

**Fig. 1.** Cross-sectional schematic of the surface-illuminated lateral Ge-on-Si SACM APD.

Silicon photonics is attractive for these applications because it combines high-speed optical functions with CMOS-compatible fabrication and scalable integration [3]. Although 1550 nm is the standard wavelength for fiber-optic communication, it is also widely used in LiDAR, free-space optical links, short-wave infrared imaging, remote sensing, spectroscopy, and industrial monitoring [4], [5]. Operation near 1550 nm also benefits from established C-band lasers, fibers, and optical components, as well as favorable eye-safety limits for free-space systems. These advantages motivate compact, sensitive, and high-speed 1550-nm photodetectors that can be integrated on Si.

Silicon has negligible interband absorption at 1550 nm because of its large bandgap. Ge is therefore a practical absorber for Si photonics at 1550 nm because it provides light absorption at this wavelength and remains compatible with CMOS processing [6], [7], in the form of Ge-on-Si photodetectors. For applications requiring high responsivity, a photodetector with internal gain is desirable. Avalanche photodetectors (APDs) are generally used for high-gain detection. Among all designs, Ge/Si separate-absorption-charge-multiplication (SACM) APDs provide an effective approach by spatially separating optical absorption from avalanche multiplication [8]. Photocarriers are generated in the Ge absorption region and then transported to a high-field Si multiplication region. This architecture combines the band-edge absorption of Ge at 1550 nm with the favorable impact-ionization characteristics of Si, enabling enhanced photoresponse [6], [7], [9]. Most high-speed Ge/Si APDs for optical interconnects use waveguide coupling because it provides strong optical confinement and short carrier-transport distances [9], [10].

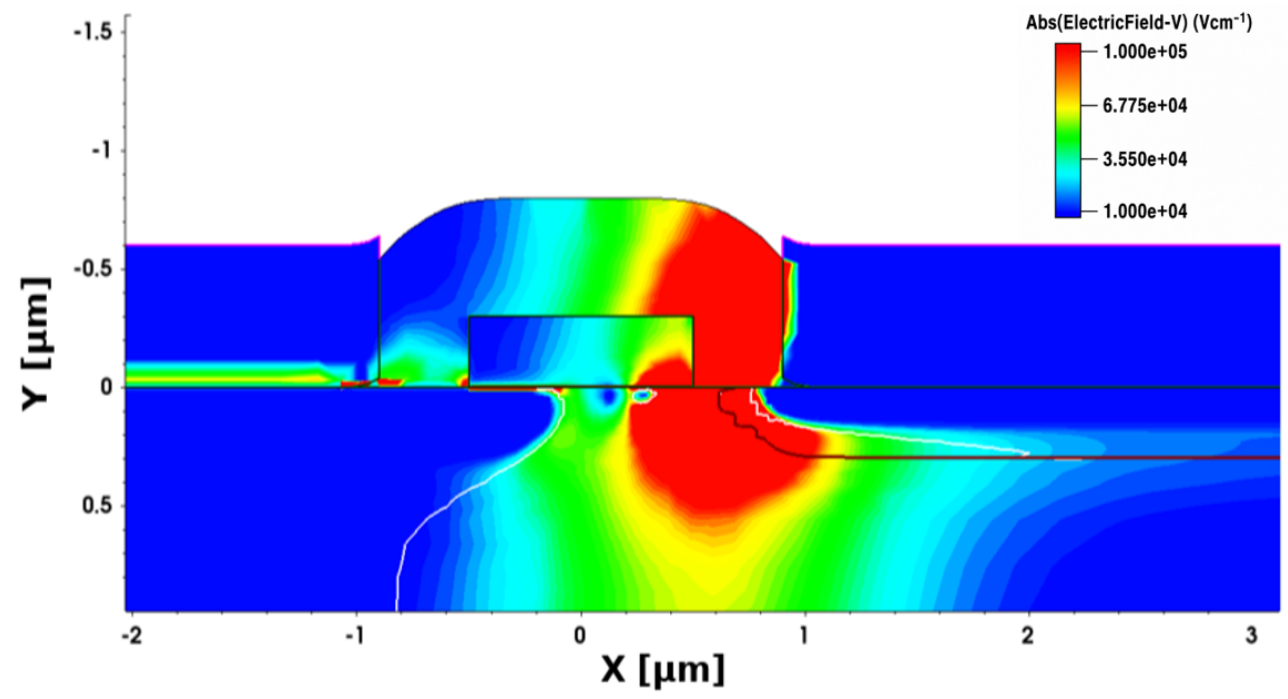

**Fig. 2.** TCAD electric-field distribution of the Ge-on-Si APD at −15 V showing field suppression beneath the Ge absorber and field localization toward the $N^+$-side Si multiplication region.

However, waveguide-coupled devices require input waveguides and optical couplers, which makes them less suitable for direct free-space illumination. Surface illumination is attractive for LiDAR, imaging, sensing, and detector arrays because light can enter the detector surface normally without on-chip waveguide routing. It also supports direct pixel access and can simplify one- and two-dimensional detector arrays.

Several Ge-on-Si devices have demonstrated surface or vertical illumination. Kim et al. reported a vertical-illumination Ge-on-Si APD with high-speed operation [11]. Li et al. demonstrated a surface-illuminated interdigitated Ge-on-Si detector at 1550 nm with high responsivity and direct optical access, but limited bandwidth and modest gain [12]. Hsu et al. later introduced substrate-embedded lateral Si p–i–n junctions to preserve a metal-free optical aperture and reduce dark current, and suggested that transit time mainly limited the high-speed response [13]. Fleming et al. reported a surface-normal pseudo-planar Ge-on-Si APD with high avalanche gain and low excess noise [14], and Modak et al. extended the concept to a CMOS-compatible 10-pixel Ge-on-Si APD array [4]. These studies show that surface illumination is attractive for free-space and arrayed receivers, but simultaneously achieving high responsivity and high bandwidth remains difficult because carrier-transport distance, capacitance, and electric-field distribution are strongly coupled.

This work develops a surface-illuminated lateral interdigitated Ge-on-Si SACM APD that separates Ge absorption from Si avalanche multiplication through lateral electric-field engineering. The Ge mesa is illuminated directly from the surface and serves as the absorption region, while multiplication occurs mainly in the laterally patterned Si region. A $P^-$ charge-control region and interdigitated $P^+/N^+$ Si regions redistribute the lateral potential, suppress high-field penetration toward the Ge absorber and Ge/Si heterointerface, and localize the strongest field near the $N^+$-side Si multiplication region. This design mitigates local heterointerface field crowding while preserving direct surface-normal optical access. To establish the design rules, we systematically varied the intrinsic-region spacing, $P^-$ charge-control width, and interdigitated finger geometry and correlated them with breakdown voltage, dark current, avalanche gain, responsivity, and RF bandwidth. Figure 1 displays the proposed device structure.

## II. DEVICE DESIGN AND FABRICATION

We first modeled the fabrication sequence in Sentaurus Process, then imported the resulting structures into Sentaurus Device for electrical and optoelectronic simulation. The device model included doping-dependent mobility, high-field velocity saturation, interface effects, Shockley–Read–Hall recombination, Auger recombination, Hurkx tunneling, and band-to-band tunneling. We calculated avalanche multiplication with the Van Overstraeten–de Man model, in which the local electron and hole impact-ionization coefficients follow Chynoweth-type field dependence [15].

$$\gamma(T) = \frac{\tanh\left(\frac{\hbar\omega_{op}}{2k_B T_0}\right)}{\tanh\left(\frac{\hbar\omega_{op}}{2k_B T}\right)} \quad (1)$$

$$\alpha(F,T) = \gamma(T)a\,\exp\left(-\frac{\gamma(T)b}{F}\right) \quad (2)$$

where $\alpha$ is the impact-ionization coefficient, $F$ is the avalanche driving force, and $\gamma(T)$ is the temperature correction term, incorporating optical phonon energy $\hbar\omega_{op}$, Boltzmann constant $k_B$, 300 K reference temperature $T_0$, and simulation temperature $T$. Material fitting parameters $a$ and $b$ were provided by Sentaurus Device in addition to $\hbar\omega_{op}$. The $T$ was only used for internal model temperature corrections because all simulations were conducted at 300 K. The avalanche driving force, $F$, was obtained based on carriers' quasi-Fermi level gradients. The doping profile was designed to keep the field relatively low beneath the Ge absorber and near the Ge/Si heterointerface, thereby suppressing unwanted field penetration and impact ionization in Ge while maintaining efficient carrier extraction [6]. As shown in Fig. 2, the TCAD simulation reveals a strongly nonuniform lateral electric-field distribution. The highest electric field is concentrated on the $N^+$-side Si region, below and adjacent to the Ge mesa, while the field beneath most of the Ge absorber remains substantially lower. The $P^-$ charge-control region redistributes the applied potential, shifting the high-field region away from the Ge/Si heterointerface and toward the Si multiplication side. This field profile is consistent with the intended SACM operation: photocarriers are generated in the Ge absorber, transported laterally toward the $N^+$-side Si region, and undergo impact ionization mainly in Si.

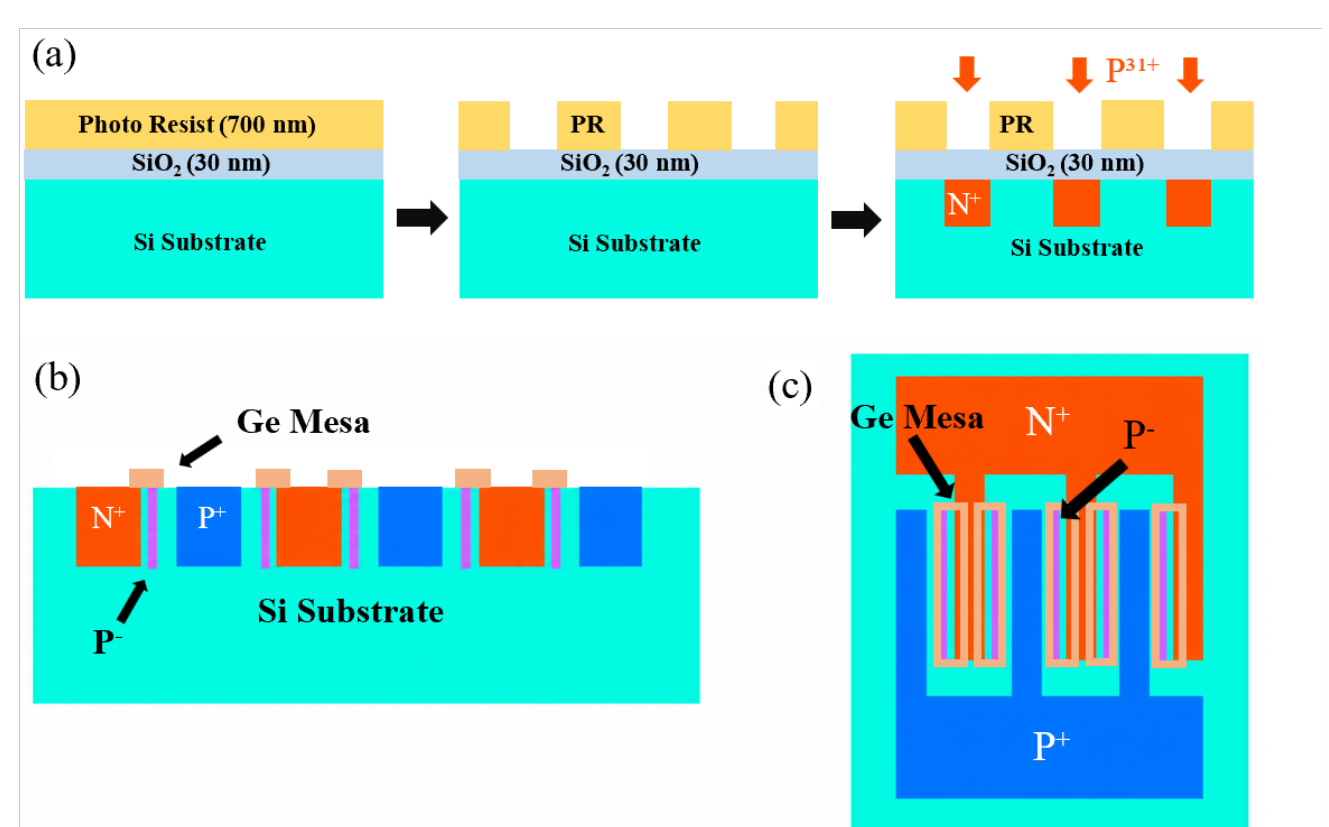

**Fig. 3.** (a) $P^{31+}$ ion implantation process flow. (b) Cross-sectional view and (c) top view of the interdigitated doped regions of the 3 × 10 layout.

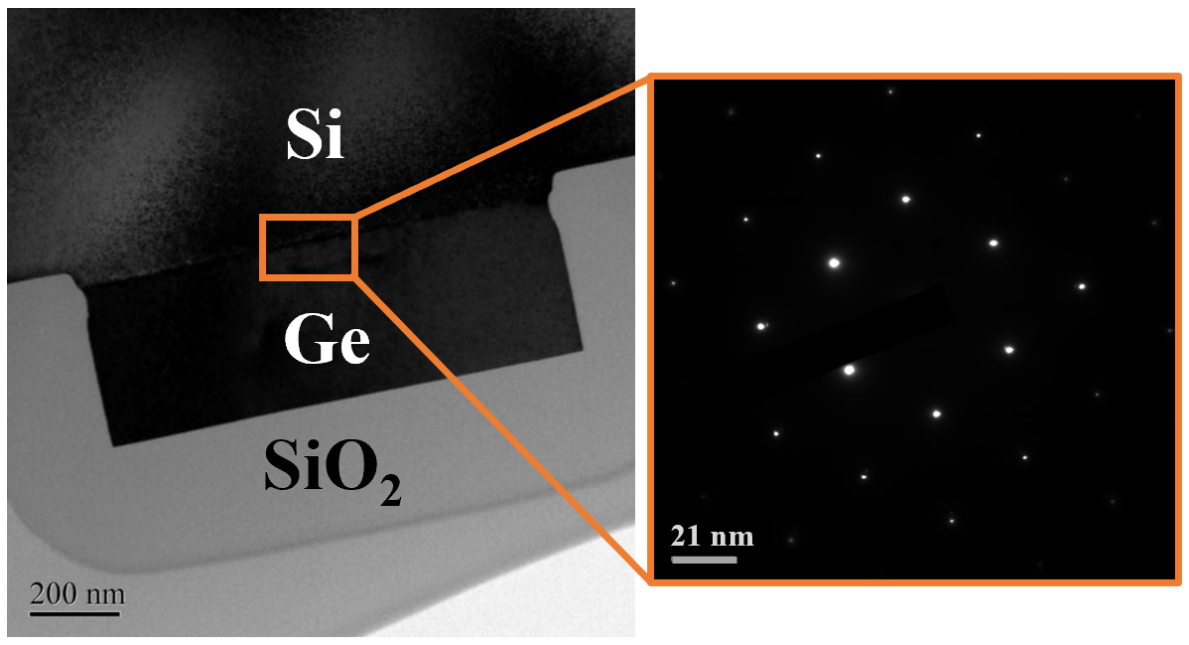

**Fig. 4.** TEM cross section of the Ge/Si interface with an electron-diffraction pattern showing aligned crystallographic orientations.

To fabricate the interdigitated Ge-on-Si APD, a 30-nm $SiO_2$ screen oxide was first grown on the Si wafer for ion implantation. The oxide reduces near-surface implantation effects, while the subsequent anneal repairs implantation damage and activates the dopants [16]. A 700-nm positive photoresist was patterned by e-beam lithography, followed by $P^{31+}$ implantation at 40 keV with a dose of $5 \times 10^{14}$ $cm^{-2}$ to form the $N^+$ regions. The $B^{11+}$ ions were then implanted at 15 keV with a dose of $5 \times 10^{15}$ $cm^{-2}$ to form the $P^+$ regions. These heavily doped regions reduce the Si/metal contact resistance [17]. The interdigitated doped regions were extended to wider contact areas so that the metal electrodes could contact Si without blocking the Ge optical aperture. The $P^-$ charge-control region was formed by $B^{11+}$ implantation at 15 keV with a dose of $8 \times 10^{12}$ $cm^{-2}$. This region shapes the lateral electric field by maintaining a relatively low field beneath the Ge absorber while concentrating the strongest field toward the $N^+$-side Si multiplication region. Rapid thermal annealing at 900 °C for 10s was then used for dopant activation and implantation-damage recovery. A 350-nm Ge absorption layer was grown by low-pressure chemical vapor deposition and patterned into mesas by e-beam lithography using a 700-nm negative photoresist.

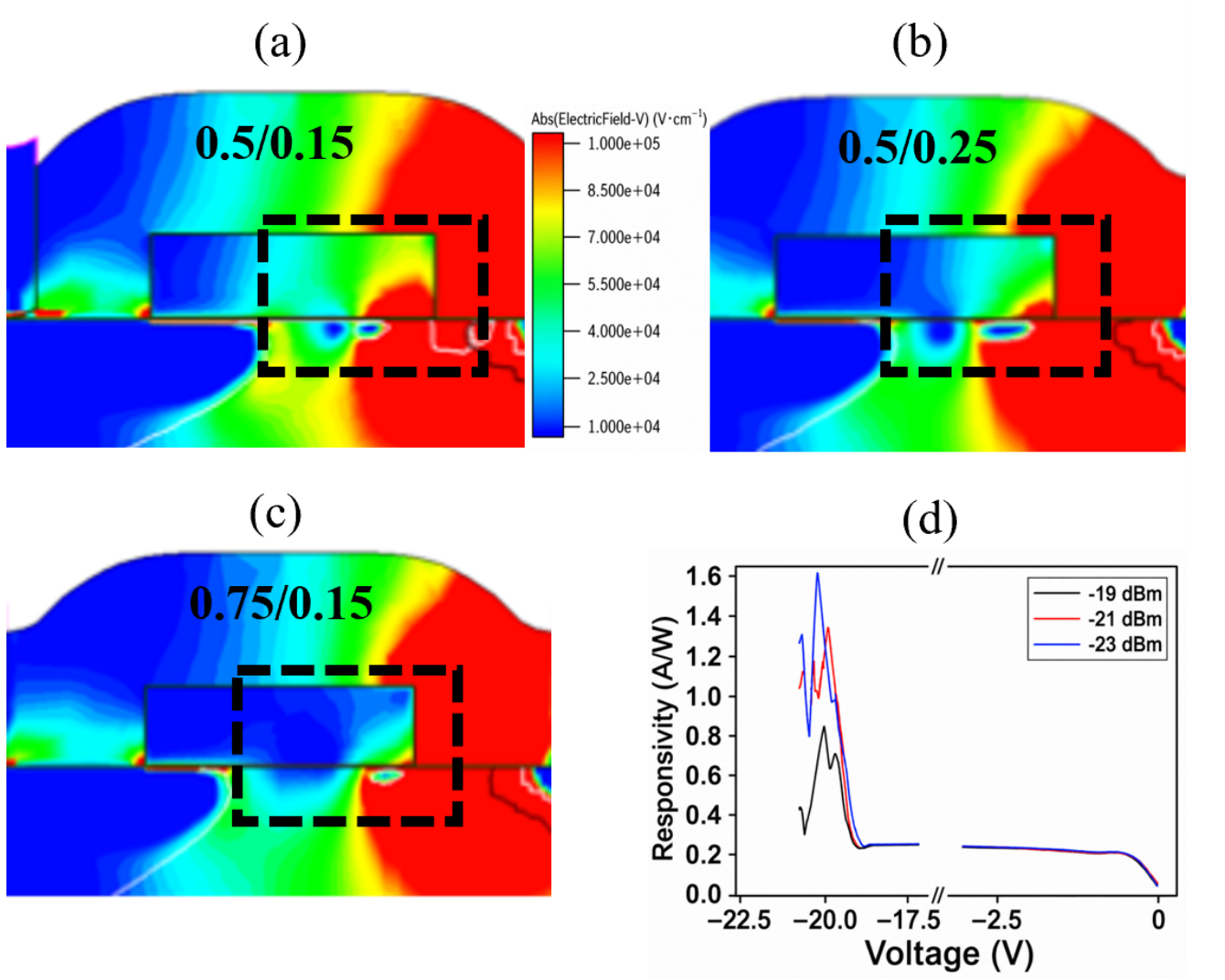

**Fig. 5.** Simulated electric-field distributions of the surface-illuminated lateral Ge-on-Si SACM APDs at −15 V for intrinsic-region/$P^-$ charge-control widths of (a) 0.5/0.15 µm, (b) 0.5/0.25 µm, and (c) 0.75/0.15 µm. (d) Simulated responsivity of the 0.75/0.15-µm device versus reverse bias at different incident optical powers.

The Ge mesa area was fixed at 30 µm² for all layouts to allow direct comparison at the same nominal absorption area. The diffraction pattern in Fig. 4 obtained from transmission electron microscopy (TEM) shows discrete spots with aligned crystallographic orientations, consistent with monocrystalline epitaxial Ge on Si. A $SiO_2$ passivation layer was deposited by plasma-enhanced chemical vapor deposition to electrically isolate non-contact regions and reduce surface-related leakage [18]. Contact openings were etched through the $SiO_2$, followed by a 9s inductively coupled plasma clean to remove residual surface oxide before metallization. The metal stack was TiN/Al–Cu/TiN/Ti from top to bottom, with the bottom Ti layer contacting Si through the openings. The electrodes were placed outside the Ge aperture to preserve direct surface-normal illumination.

Figure 5(a)–(c) shows how the lateral field profile changes with intrinsic spacing and $P^-$ width from TCAD simulation. For the intrinsic spacing/$P^-$ width equal to 0.5/0.15-µm structure [Fig. 5(a)], the field extends farther beneath the Ge mesa, although the peak remains near the $N^+$-side Si multiplication region. Increasing the $P^-$ width to 0.25 µm [Fig. 5(b)] more effectively screens the Ge absorber and shifts more of the applied potential toward the Si multiplication side. Increasing the intrinsic spacing to 0.75 µm while keeping the $P^-$ width at 0.15 µm [Fig. 5(c)] broadens the low-field transport region and increases the separation between the Ge absorber and the localized high-field Si region. These results show that the charge-control geometry mitigates local heterointerface field crowding by moving the dominant high-field region away from the Ge/Si interface and into Si. The intrinsic spacing and $P^-$ width play complementary roles in field engineering. The $P^-$ width mainly controls field screening beneath the Ge absorber and the fraction of the applied potential transferred toward the Si multiplication region.

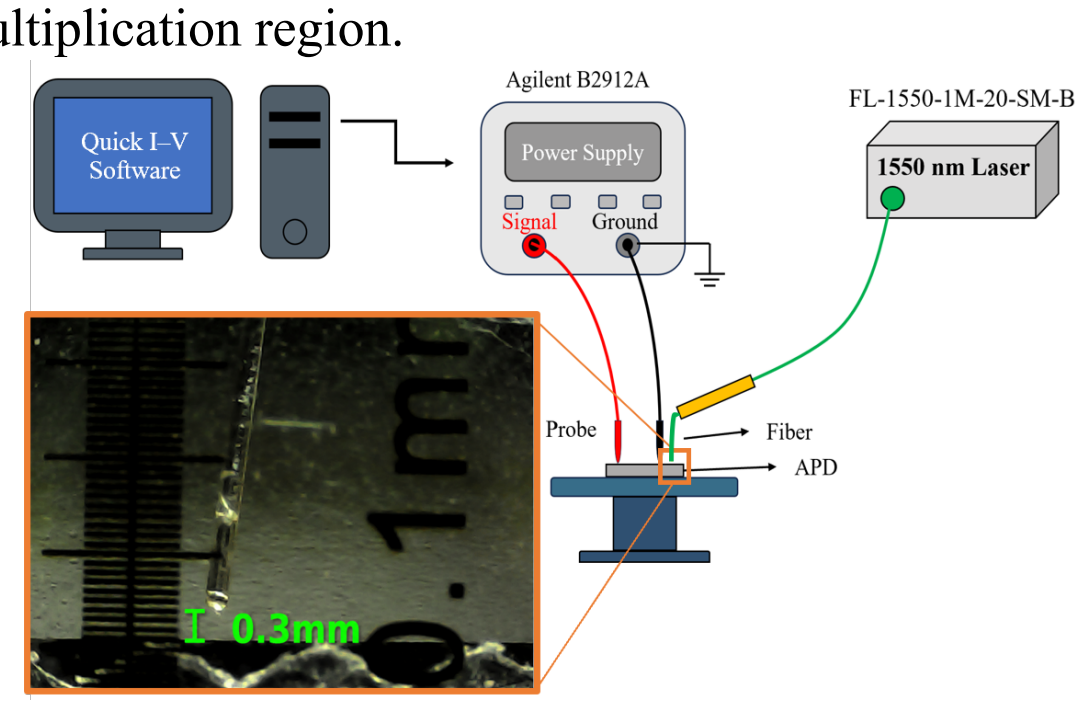

**Fig. 6.** I–V measurement setup with a magnified view of the fiber working distance.

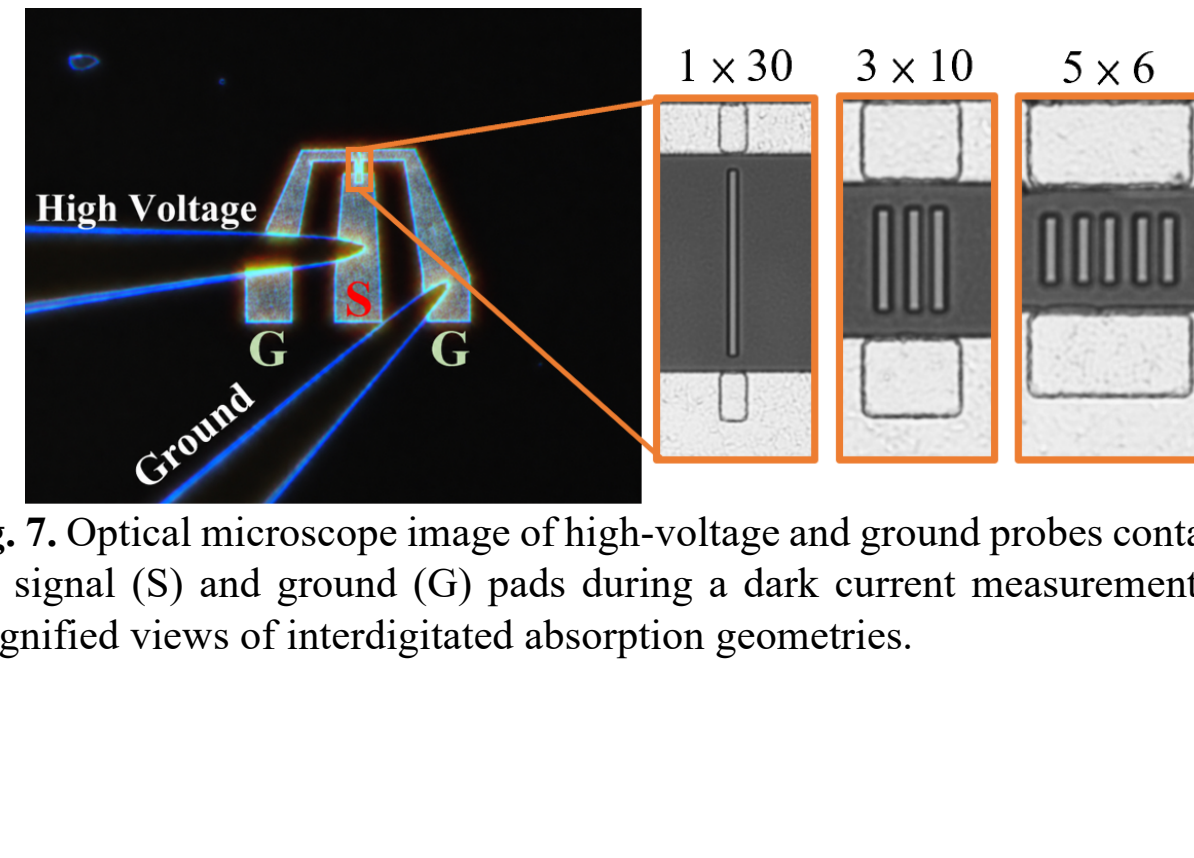

**Fig. 7.** Optical microscope image of high-voltage and ground probes contacting the signal (S) and ground (G) pads during a dark current measurement with magnified views of interdigitated absorption geometries.

The intrinsic spacing mainly controls the length of the low-field transport region and the reverse bias required to establish the high-field multiplication condition. Increasing either dimension lowers the field in the Ge-side transport region at a fixed terminal bias, suppresses leakage, and shifts punch-through and breakdown to larger reverse-bias magnitudes. Equal terminal voltages should not be interpreted as equal internal multiplication fields because the applied potential is redistributed differently among the charge-control, transport, contact, and high-field regions. Figure 5(d) also shows that responsivity is nearly independent of optical power at moderate reverse bias but becomes power dependent near avalanche. The lower responsivity at higher optical power indicates gain compression and is consistent with space-charge screening, in which increased photocarrier density reduces the effective multiplication field and impact-ionization probability [18–20].

## III. EXPERIMENTAL RESULTS

The I–V setup in Fig. 6 used 1550-nm light from a single-mode fiber with a 9-µm core positioned above the Ge absorber. Reverse-bias voltage and current were measured with a Keysight B2900A source-measure unit controlled by Quick I–V software, and tungsten probes provided electrical contact (Fig. 7). An optical microscope was used to align the fiber to the active area and verify probe contact. Measurements were first performed over a broad optical-power range, after which three input powers were selected for detailed current–voltage characterization. The photocurrent was calculated as

$$I_{ph} = I_{light} - I_{Dark}. \tag{3}$$

where $I_{light}$ and $I_{dark}$ are the currents measured with and without illumination, respectively. The fiber output power was calibrated with a Newport 918D-IR photodiode at a working distance of approximately 0.30 mm. The fiber was then moved to the APD while maintaining the same nominal working distance.

Because the illumination spot was larger than the 30 µm² Ge mesa, only part of the calibrated fiber power reached the Ge absorber. A Gaussian-beam overlap model was therefore used to distinguish the calibrated fiber power, $P_{fiber}$, from the effective power intercepted by the Ge mesa, $P_{Ge}$. The effective power was calculated as

$$P_{\text{Ge}} = \eta_c P_{\text{fiber}} \tag{4}$$

where $\eta_c$ is the calculated geometric coupling efficiency. Responsivity was then calculated using the overlap-corrected optical power as

$$R = \frac{I_{\text{ph}}}{P_{\text{Ge}}} = \frac{I_{\text{ph}}}{\eta_c P_{\text{fiber}}}. \tag{5}$$

Measurements were performed at calibrated fiber powers of 1, 1.5, and 3.75 mW. We measured seven devices for each geometry to evaluate device-to-device variation. The effect of P⁻ charge-control width was first examined by comparing 0.15- and 0.25-µm structures. As shown in Fig. 8(a), increasing the P⁻ width shifted the I–V curves toward larger reverse-bias magnitudes, consistent with stronger field screening on the Ge side and a larger voltage required to establish the avalanche-driving field in Si.

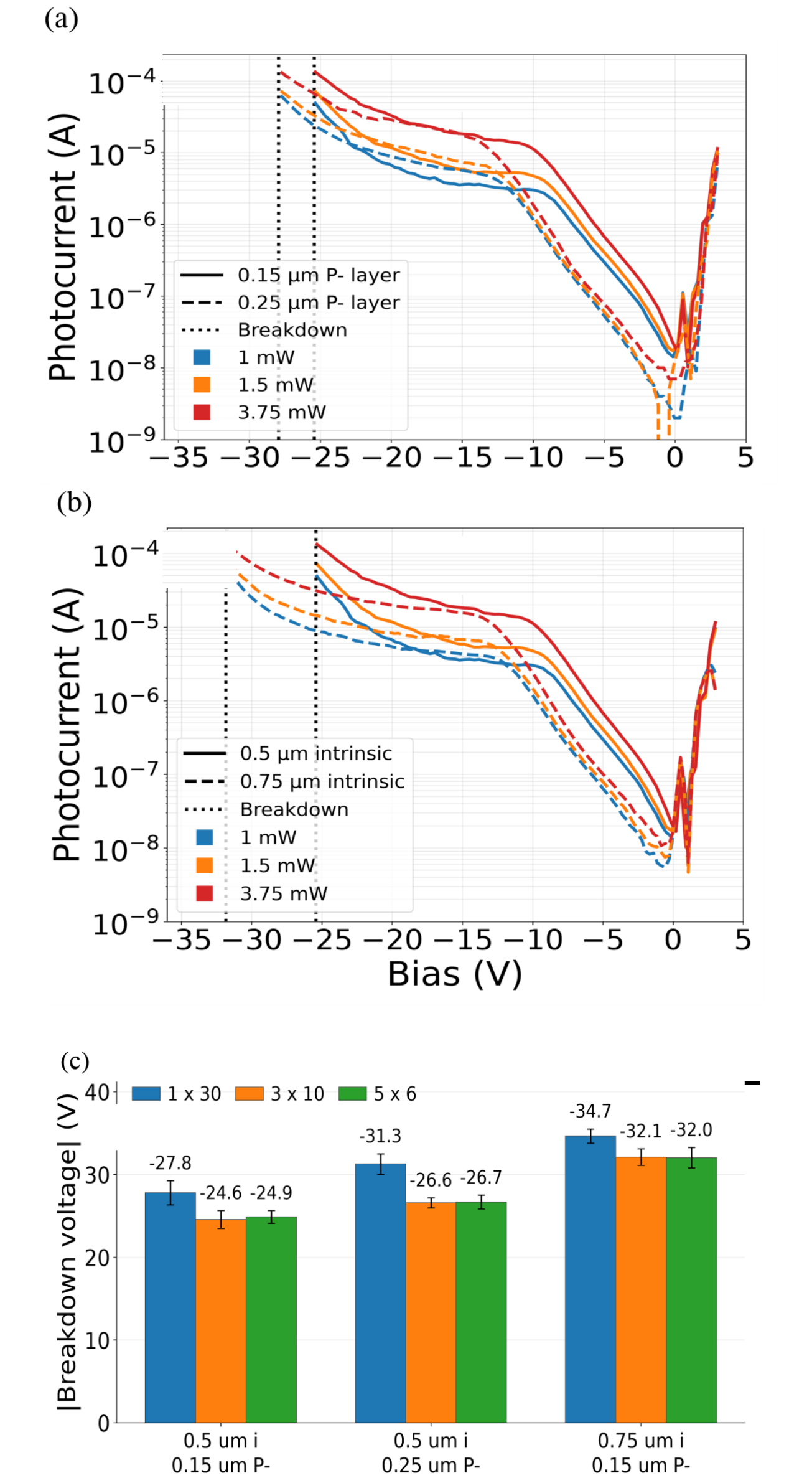

**Fig. 8.** I–V comparison for (a) P⁻ charge-control widths and (b) intrinsic-region spacings at different fiber powers. (c) Average breakdown voltage for the three interdigitated layouts.

Breakdown voltage, $V_{BR}$, was defined operationally as the bias at which the measured current exceeded 100 µA [22]. Because all layouts used the same 30 µm² Ge mesa area, this current criterion was applied consistently. For the 1 × 30 layout, increasing the P⁻ width shifted the average $V_{BR}$ by −3.5 V, compared with −2.0 and −1.8 V for the 3 × 10 and 5 × 6 layouts, respectively [Fig. 8(c)]. At −25 V, increasing the P⁻ width reduced the 1 × 30 dark current from 6.07 to 0.303 µA, a suppression of about 95%. This is a fixed-bias comparison and does not imply equal internal multiplication fields. Similarly, increasing the intrinsic spacing from 0.5 to 0.75 µm reduced the 1 × 30 dark current from 6.07 to 0.174 µA at −25 V, corresponding to about 97% suppression [Fig. 8(b)]. These results show that increasing either lateral dimension reduces the Ge-side field at a fixed terminal bias, suppresses leakage, and delays avalanche breakdown.

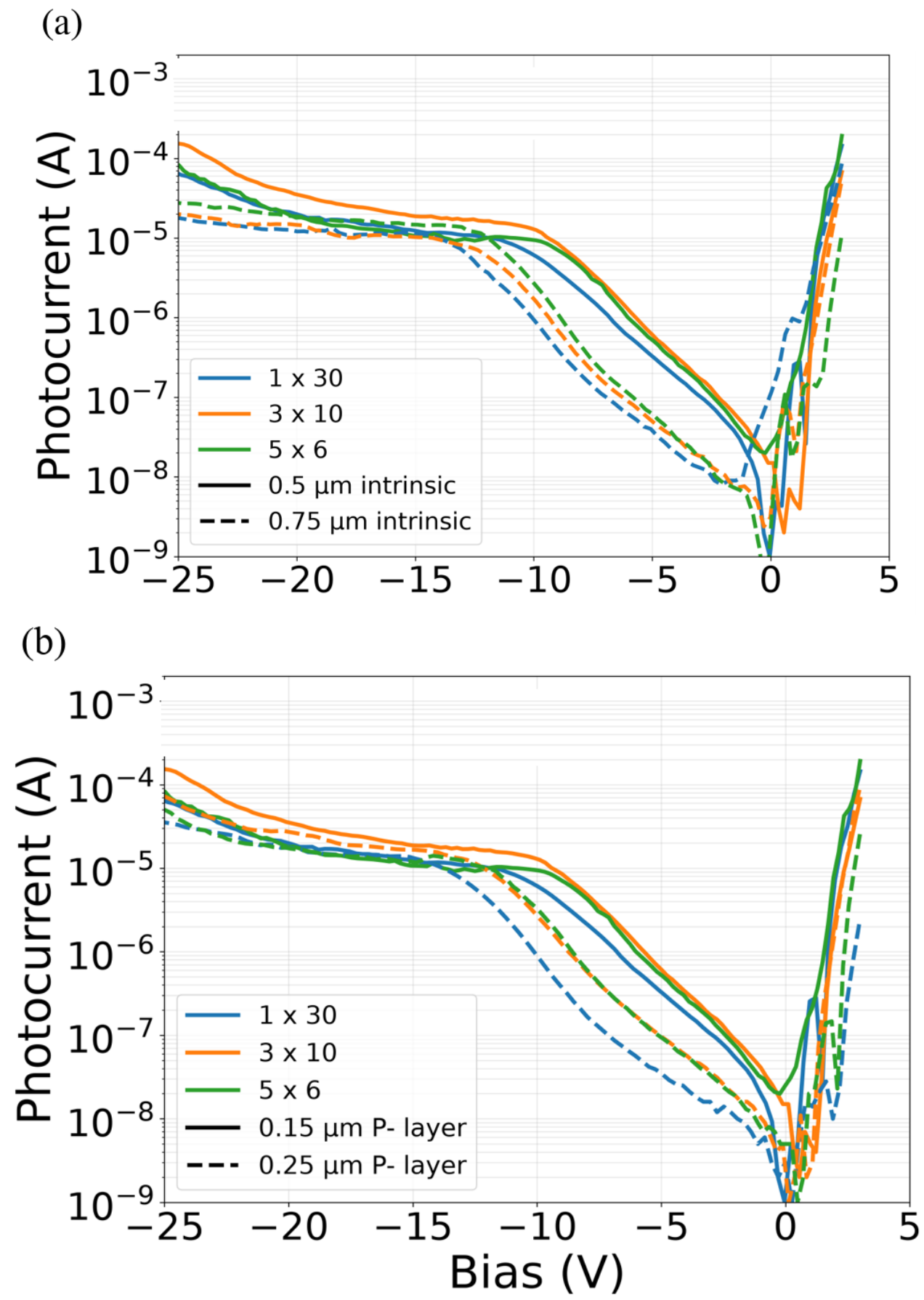

**Fig. 9.** Comparison of (a) intrinsic-region spacing and (b) $P^-$ charge-control width for different interdigitated layouts at 3.75 mW calibrated fiber power.

Because the applied voltage redistributes differently in each geometry, these dark-current reductions reflect only fixed-bias suppression, not comparisons at equal internal multiplication field.

The 3 × 10 and 5 × 6 layouts showed higher dark current and lower bias tolerance than the 1 × 30 layout. Their larger number of finger ends and greater perimeter per unit active area may increase edge-related field crowding and leakage. This explanation is consistent with the measured trend, but three-dimensional corner-field simulations would be required for direct confirmation [23], [24].

The 3 × 10 and 5 × 6 layouts showed similar illumination behavior, consistent with their similar Gaussian-beam overlap [Fig. 9]. We modeled the beam at the device plane using a 0.30-mm fiber working distance and a nominal waist radius of 5.2 µm [25]. The model assumes normal incidence and ideal lateral alignment, so uncertainty in fiber position, tilt, and beam waist contributes to the responsivity error. The calculated coupling efficiencies, $\eta_c$, were 1.930%, 2.208%, and 2.181% for the 1 × 30, 3 × 10, and 5 × 6 layouts, respectively. The lower coupling efficiency of the 1 × 30 layout must therefore be considered when comparing raw photocurrent, while responsivity is calculated from the overlap-corrected power intercepted by the Ge mesa.

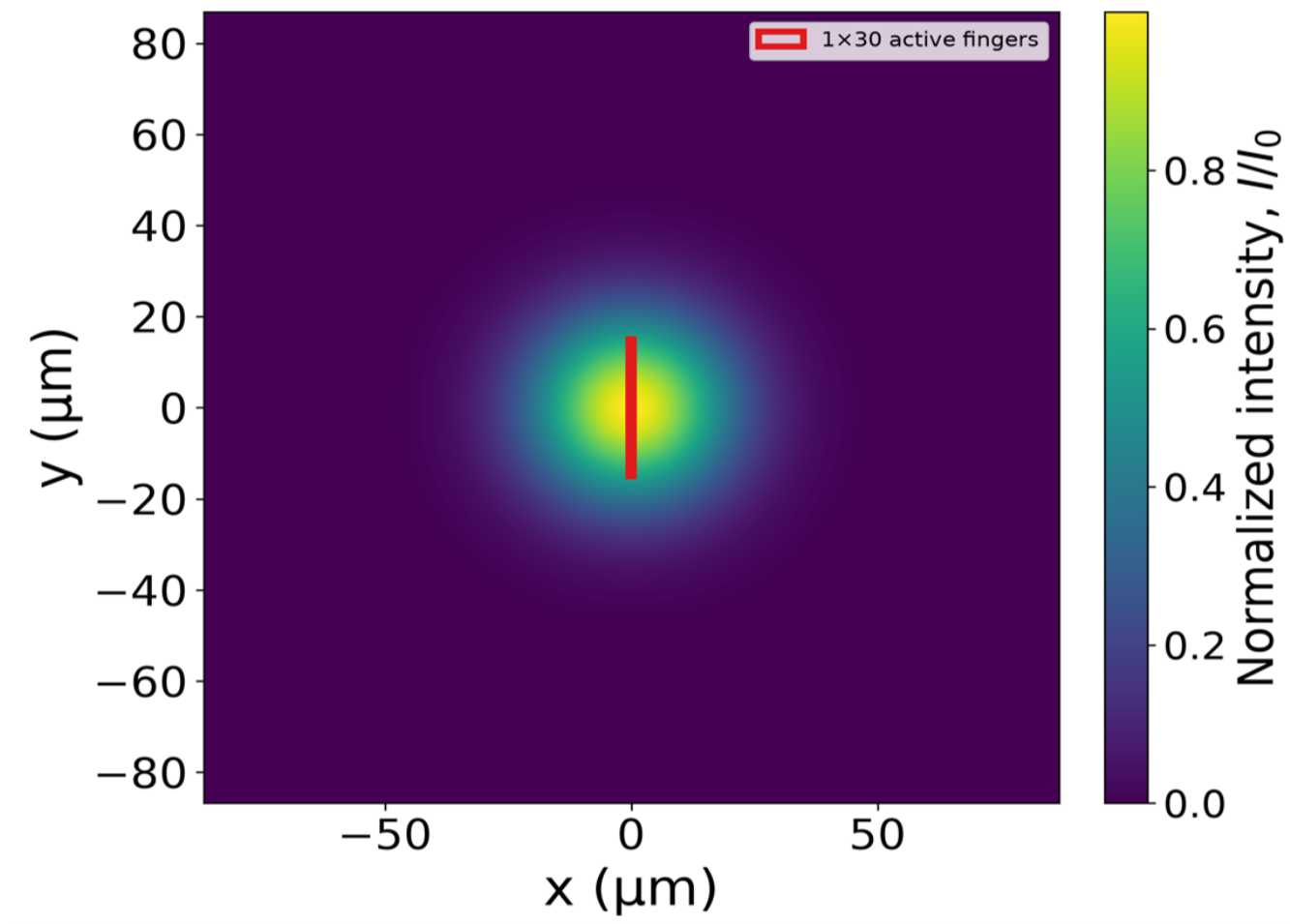

**Fig. 10.** Gaussian-beam intensity distribution and overlap with the 1 × 30 Ge mesa. The calculated geometric coupling efficiency is $\eta_c$ = 1.930%.

The punch-through voltage, $V_{PT}$, is the bias at which the depletion region extends far enough through the absorption and transport regions to enable efficient photocarrier extraction. It is typically followed by a near-unity-gain region before strong avalanche multiplication [26], [27]. In this work, $V_{PT}$ was identified from the maximum slope of the photocurrent–voltage curve [27].

The unity-gain reference voltage, $V_{M=1}$, was identified in the post-punch-through region from the second derivative of photocurrent with respect to voltage [27]. When no clear plateau was present, a local minimum in the second derivative was used as an operational estimate. Because derivative-based extraction is sensitive to measurement noise and smoothing, the extracted unity-gain responsivity and multiplication gain are treated as operational estimates rather than exact field markers. Gain was calculated separately at each optical power, and geometry comparisons used the lowest optical power to reduce space-charge-screening effects.

$$M(V) = \frac{I_{ph}(V)}{I_{ph}(V_{M=1})} \tag{7}$$

**TABLE I**
**PERFORMANCE COMPARISON OF INTERDIGITATED FINGER LAYOUTS**

| Layout | Avg. gain at 0.9 $V_{BR}$ | Max. gain | Avg. R at 0.9 $V_{BR}$ (A/W) | Peak R (A/W) | Avg. $V_{PT}$ (V) | Avg. $V_{M=1}$ (V) |
|---|---|---|---|---|---|---|
| 1 × 30 | 10.71 ± 3.46 | 124.94 | 1.03 ± 0.42 | 11.81 | −10.73 ± 1.61 | −12.04 ± 2.29 |
| 3 × 10 | 8.37 ± 3.46 | 54.33 | 0.98 ± 0.48 | 4.61 | −9.94 ± 1.53 | −11.59 ± 3.18 |
| 5 × 6 | 7.52 ± 3.16 | 88.38 | 0.95 ± 0.48 | 5.55 | −9.40 ± 1.53 | −10.50 ± 3.18 |

TABLE II
EFFECT OF INTRINSIC AND P⁻ CHARGE-CONTROL WIDTHS FOR 1 × 30 DEVICES

| Intrinsic/P⁻ width (µm) | Gain at 0.9 $V_{BR}$ (range) | Max. gain | Avg. R at 0.9 $V_{BR}$ (A/W) | Peak R (A/W) | Peak-bias range (V) |
|---|---|---|---|---|---|
| 0.5/0.15 | 11.66–19.06 | 124.94 | 1.10 ± 0.39 | 11.81 | −28.68 to −24.54 |
| 0.5/0.25 | 4.50–11.55 | 57.60 | 1.07 ± 0.57 | 4.75 | −33.23 to −28.92 |
| 0.75/0.15 | 8.11–13.03 | 56.87 | 0.92 ± 0.29 | 4.99 | −36.00 to −33.05 |

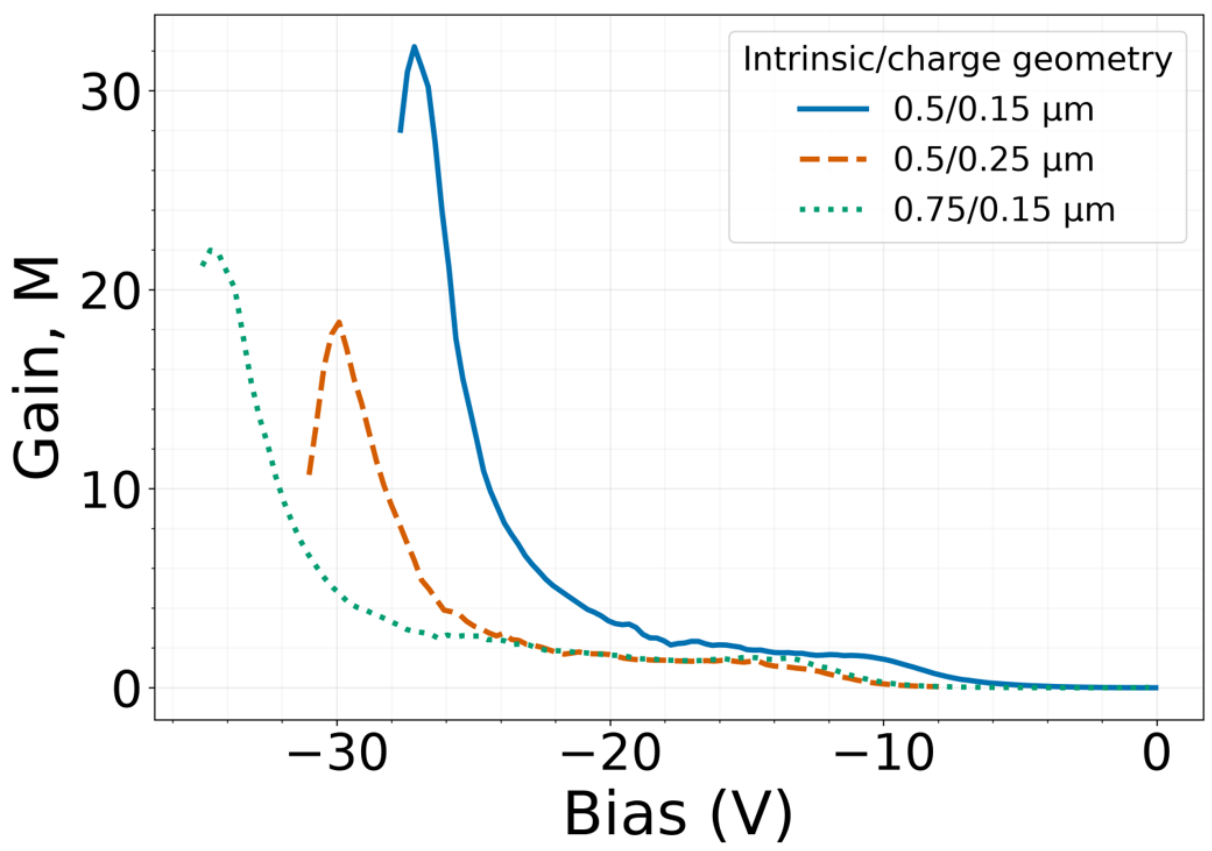

**Fig. 11.** Avalanche-gain curves of the 1 × 30 layout for different intrinsic-region spacings and P⁻ charge-control widths.

Average gain and responsivity were compared at 0.9 $V_{BR}$ rather than at a single terminal bias so that devices with different breakdown voltages were evaluated at comparable positions within their stable operating ranges. This normalization provides a consistent comparison of operating regimes but does not imply identical internal electric-field distributions.

As the number of interdigitated fingers increased, the average gain and responsivity at 0.9 $V_{BR}$ decreased, while the punch-through and unity-gain voltages shifted to smaller reverse-bias magnitudes (Table I). The higher edge density of the multi-finger layouts may increase local field crowding and leakage, reducing the fraction of applied bias that contributes to useful avalanche multiplication. This interpretation is consistent with the dark-current trend but requires three-dimensional field mapping for direct confirmation [23], [24].

The 1 × 30 layout was selected for further comparison because it provided the best overall balance of dark current, responsivity, and bandwidth. Within this layout, the 0.5/0.15-µm structure showed the highest measured gain and responsivity within the stable bias range. As summarized in Table II, increasing either the intrinsic spacing or the P⁻ charge-control width shifted avalanche operation to larger reverse-bias magnitudes and reduced the measured gain and responsivity before breakdown.

These results should not be interpreted as evidence that a longer lateral region intrinsically produces lower gain at the same internal multiplication field. The internal field was not measured directly, and the applied voltage is redistributed across several lateral regions. TCAD instead indicates that the larger structures broaden the low-field transport region and reduce the field beneath the Ge mesa at a given terminal bias. A larger reverse bias is therefore required to establish a comparable ionization integral near the N⁺-side high-field region. At 0.9 $V_{BR}$, increasing the P⁻ width produced a larger reduction in gain and responsivity than increasing the intrinsic spacing by 0.25 µm (Fig. 11), indicating that the P⁻ layer more strongly modifies the lateral field and carrier injection into the Si multiplication region within the investigated design range. Avalanche gain depends on the accumulated impact-ionization probability along the high-field region rather than on peak field or geometrical length alone. The electron and hole ionization integrals are

$$I_e = \int_0^{W_{\text{eff}}} \alpha\,[E(x)]\,dx \tag{8}$$

$$I_h = \int_0^{W_{\text{eff}}} \beta\,[E(x)]\,dx \tag{9}$$

where α[E(x)] and β[E(x)] are the local electron and hole impact-ionization coefficients, respectively, and $W_{\text{eff}}$ is the effective high-field carrier-transport region. Increasing the lateral dimensions redistributes the applied voltage, broadens the low-field transport region, and requires a larger reverse bias to establish comparable ionization conditions near the N⁺-side Si region.

Devices with different geometries can therefore show similar gain when their integrated ionization probabilities are comparable, even if their terminal voltages or peak electric fields differ. Figure 12 also shows that avalanche-enhanced responsivity decreases as optical power increases, providing experimental evidence of gain compression in the high-field regime. Together with the power-dependent TCAD result in Fig. 5(d), this behavior is consistent with space-charge screening: photogenerated charge partially reduces the multiplication field and lowers the effective impact-ionization probability. Because the internal field under illumination was not measured directly, we interpret the result as space-charge-induced gain compression rather than as a quantitative isolation of the screening mechanism. TCAD pulse-response simulations further predicted higher 3-dB bandwidth for devices with smaller intrinsic spacing and P⁻ width.

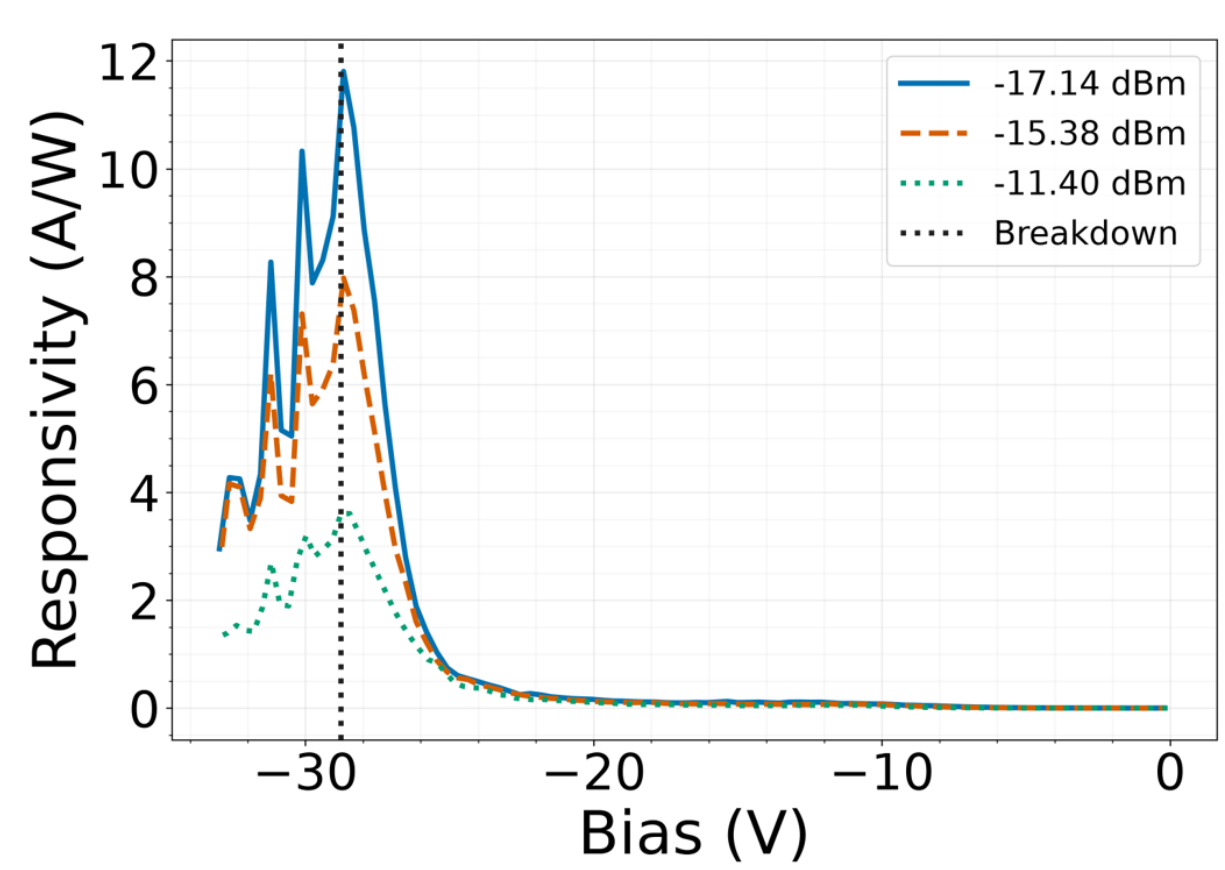

**Fig. 12.** Avalanche-enhanced responsivity versus bias for the 1 × 30, 0.5/0.15-µm device at different effective received optical powers.

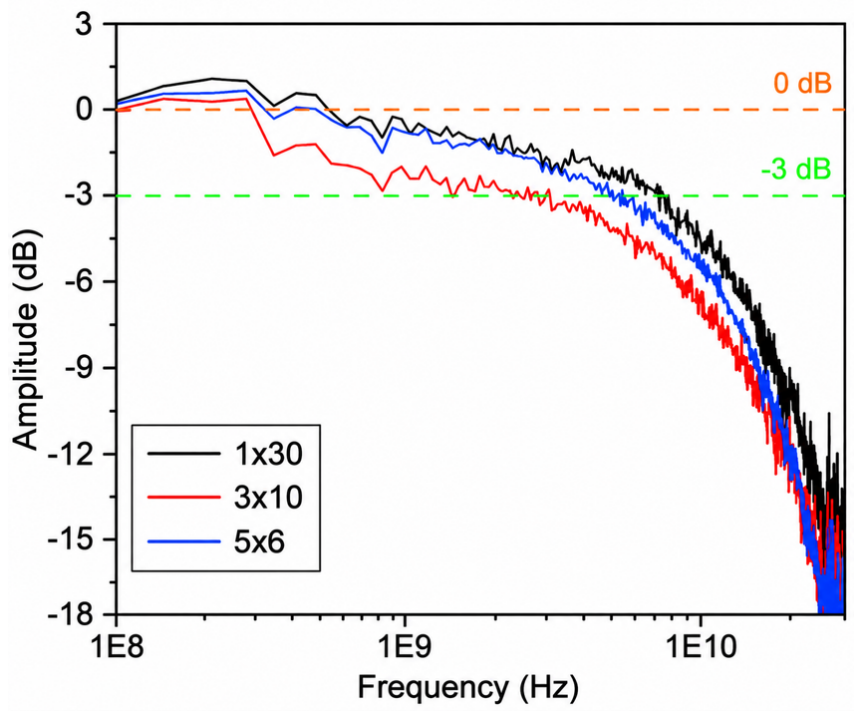

**Fig. 13.** Normalized small-signal frequency responses of the 1 × 30, 3 × 10, and 5 × 6 layouts for the 0.5/0.15-µm geometry at −20 V at 1550 nm. The measured 3-dB bandwidths are 7.13, 1.41, and 4.88 GHz, respectively.

This trend is consistent with a shorter low-field transport path and faster delivery of photocarriers to the localized high-field Si region. The 0.5/0.15-µm geometry was therefore selected for RF characterization of the three interdigitated layouts (Table III). Frequency responses were measured from 10 MHz to 67 GHz at −20 V using modulated 1550-nm optical sources. At 1550 nm, the corresponding bandwidths were 7.13, 1.41, and 4.88 GHz. The 1 × 30 device provided the highest matched-bias GBP of 27.07 GHz at 1550 nm, while the 3 × 10 and 5 × 6 devices showed higher gain but lower bandwidth at the same bias, illustrating the gain–speed trade-off. The gain-bandwidth product is defined as

$$\mathrm{GBP} = M f_{3\mathrm{dB}}. \tag{10}$$

The $S_{21}$ response was measured with a Keysight N5227B vector network analyzer calibrated to the cable end. A bias tee and GSG probe supplied the −20-V reverse bias and collected the RF response. No RF amplifier, equalizer, or digital signal processing was used. The 1550-nm lasers were coupled through a Keysight N4373E Lightwave Component Analyzer with a 3-dBm output. Each response was normalized to its 10-MHz value, and the 3-dB bandwidth was obtained from the fitted response at the frequency 3 dB below this reference. GBP values were calculated using gain and bandwidth measured at the same −20-V bias under 1550-nm illumination.

**TABLE III**
**RF PERFORMANCE OF 0.5/0.15-µm DEVICES AT −20 V**

| Finger layout | Gain at −20 V | 3-dB BW, 1550 nm (GHz) | GBP, 1550 nm (GHz) |
|---|---|---|---|
| 1 × 30 | 3.796 | 7.13 | 27.07 |
| 3 × 10 | 5.014 | 1.41 | 7.07 |
| 5 × 6 | 5.517 | 4.88 | 26.92 |

A complementary low-voltage Ge p–i–n detector based on doping-engineered carrier confinement recently achieved 370 A/W responsivity at −0.5 V but only about 148.8 MHz measured bandwidth [28]. This result illustrates the strong gain–speed trade-off associated with long carrier residence time. In contrast, the present SACM APD uses impact-ionization multiplication in Si while maintaining GHz-class bandwidth.

**TABLE IV**
**COMPARISON OF REPRESENTATIVE Ge/Si AVALANCHE PHOTODETECTORS**

| Ref. | Architecture | λ (nm) | $R_0$ (A/W) | M | BW (GHz) | GBP (GHz) |
|---|---|---|---|---|---|---|
| **[4]** | Surface array | 1550 | — | 20 | — | — |
| **[9]** | Waveguide lateral | 1550 | 0.78 ± 0.1 | — | 27 | — |
| **[11]** | Vertical SACM | 1550 | — | 15.33 | 30 | 460 |
| **[12]** | Surface interdigitated | 1550 | — | ~2 | $8.43 \times 10^{-5}$ | $1.69 \times 10^{-4}$ |
| **[14]** | Surface pseudo-planar | 1550 | 0.41 | 101 (max.) | — | — |
| **[29]** | Waveguide reach-through | 1310 | 0.93 | 12.8 | 48 | 615 |
| **This work[a]** | Surface lateral SACM | 1550 | 1.10 ± 0.39 | 3.796 | 7.13 | 27.07 |

[a] For this work, M, BW, and GBP were measured at the same −20 V bias.

Table IV compares representative surface-illuminated and waveguide-coupled Ge/Si APDs using only experimentally reported metrics. An em dash indicates a value that was not explicitly reported. Waveguide-coupled devices generally achieve higher bandwidth and GBP because of strong optical confinement and short carrier-transport distances. Surface-normal devices instead offer direct optical access and easier integration with free-space or arrayed detection. The present device lies between these regimes, providing a 7.13-GHz bandwidth while retaining direct surface-normal access and a lateral, Si-dominated multiplication region.

The 27.07-GHz GBP reported here is lower than that of optimized waveguide Ge/Si APDs, such as the 615-GHz reach-through device of Xiang et al. [29]. The main contribution is therefore not record GBP, but the combination of GHz-class response, direct surface-normal access, separated Ge absorption and Si multiplication, and systematic lateral electric-field engineering. Compared with the surface-illuminated interdigitated detector of Li et al. [12], the present device provides substantially higher measured bandwidth while preserving the interdigitated surface-illumination concept. Fleming et al. [14] and Modak et al. [4] demonstrate complementary strengths in excess-noise characterization and array implementation. The present architecture is therefore best viewed as a complementary solution for free-space and arrayed 1550-nm detection rather than a replacement for high-GBP waveguide APDs.

## IV. CONCLUSION

Surface-illuminated lateral Ge-on-Si SACM APDs with different intrinsic spacings, $P^-$ charge-control widths, and interdigitated layouts were simulated and experimentally evaluated. TCAD indicates that Ge provides the 1550-nm absorption region while avalanche multiplication occurs mainly in Si. The $P^-$ charge-control layer redistributes the lateral field, suppresses high-field penetration toward the Ge/Si heterointerface, and concentrates the high-field region in Si. Increasing either the intrinsic spacing or $P^-$ width reduced dark current at a fixed bias and shifted punch-through, unity gain, and breakdown to larger reverse-bias magnitudes. Among the tested structures, the 1 × 30 layout with 0.5/0.15-µm intrinsic/$P^-$ dimensions provided the best overall performance, reaching a peak responsivity of 11.81 A/W. At −20 V, it achieved a gain of 3.796, a 7.13-GHz bandwidth at 1550 nm, and a 27.07-GHz GBP. TEM and electron diffraction support crystalline registry at the examined Ge/Si interface, while the optical-power dependence provides evidence of avalanche-gain compression consistent with space-charge screening. Overall, the results show that lateral charge-control engineering can combine heterointerface field management, dark-current suppression, avalanche gain, and GHz-class response in a surface-illuminated Ge-on-Si APD.